\pdfoutput=1
\documentclass[useAMS,usenatbib]{mn2e}
\usepackage[pdftex]{graphicx}
\usepackage{times}
\usepackage{wasysym}
\usepackage{url}
\usepackage{lineno}
\usepackage{amssymb}

\newcommand{\ngc}{NGC\,253}
\newcommand{\g}{$\gamma$}

\newcommand{\UNITS}[1]{\,\mathrm{#1}}

\title{Non-thermal emission from Pulsar-Wind Nebulae in Starburst
  Galaxies}

\author[S. Ohm and J.A. Hinton]{\normalsize S.~Ohm,$^{1,2}$ and
  J.A.~Hinton$^{1}$
  \\
  $^{1}$
  Department of Physics and Astronomy, The University of Leicester, University Road, Leicester, LE1 7RH, United Kingdom \\
  $^{2}$
  School of Physics \& Astronomy, University of Leeds, Leeds LS2 9JT, UK \\
}

\begin{document}

\date{Accepted 2012 October 31. Received 2012 October 23; in original form
  2012 September 21}
\pagerange{\pageref{firstpage}--\pageref{lastpage}} \pubyear{2012}

\maketitle

\label{firstpage}

\begin{abstract}
  The recently detected \g-ray emission from Starburst galaxies is
  most commonly considered to be diffuse emission arising from strong
  interactions of accelerated cosmic rays. \citet{SB:Mannheim12},
  however, have argued that a population of individual pulsar-wind
  nebulae (PWNe) could be responsible for the detected TeV
  emission. Here we show that the Starburst environment plays a
  critical role in the TeV emission from Starburst PWN, and perform
  the first detailed calculations for this scenario. Our approach is
  based on the measured star-formation rates in the Starburst nuclei
  of \ngc\ and M\,82, assumed pulsar birth periods and a simple model
  for the injection of non-thermal particles. The two-zone model
  applied here takes into account the high far-infrared radiation
  field, and different densities and magnetic fields in the PWN and
  the Starburst regions, as well as particle escape. We confirm that
  PWN can make a significant contribution to the TeV fluxes, provided
  that the injection spectrum of particles is rather hard and that the
  average pulsar birth period is rather short ($\sim35$ ms). The PWN
  contribution should lead to a distinct spectral feature which can be
  probed by future instruments such as CTA.
\end{abstract}


\maketitle

\section{Introduction}\label{sec:intro}

Starburst (SB) galaxies have a very high star-formation rate (SFR) and
hence supernova (SN) rate in a very localised region -- the SB region
-- which is often located in the centre of the galaxy. The paradigm of
cosmic-ray acceleration in supernova remnant shock fronts and the very
large amounts of gas in the SB region, available for proton-proton
interactions and subsequent $\pi^0$-decay \g-ray emission, made them
suspected emitters of high-energy (HE; $100\,\UNITS{MeV} \leq E \leq
100\,\UNITS{GeV}$) and very-high-energy (VHE; $E\ge100\,\UNITS{GeV}$)
\g-rays\citep[e.g.][]{Voelk89,M82:Akyuz91} emission. Indeed, \g-ray
emission from the prototypical SB galaxies \ngc\ and M\,82 using the
space-based Fermi-LAT \citep{Fermi:NGC253M82} and the ground-based
Cherenkov telescopes H.E.S.S. and VERITAS \citep{HESS:NGC253,
  VERITAS:M82} has recently been reported. There are no indications of
\g-ray variability, and in the one case where the instrumental
resolution is sufficient to discriminate, the TeV signal from \ngc,
the emission is compatible with originating in the SB nucleus, without
a significant contribution from the disc of the galaxy
\citep{HESS:NGC253_2012}.

Interpretations of the detected GeV and TeV \g-ray emission from SB
galaxies are possible within the traditional framework if $\sim10$\%
of the kinetic energy of SN explosions is converted into particle
acceleration and about $(20 - 30)$\% of these protons lose their
energy in interactions with target material \cite[see
e.g.][]{SB:Lacki11,OhmHinton12,Paglione12,HESS:NGC253_2012}. Whilst
this scenario is certainly plausible, it may be somewhat
oversimplified. For example, it is not clear that a contribution from
individual sources can safely be neglected at all energies. The TeV
\g-ray source population observed in our own Galaxy is diverse and
shows emission from \g-ray binaries, stellar clusters, pulsar wind
nebulae (PWNe), as well as SNR, with PWNe being the dominant object
class \citep[e.g.][]{HintonHofmann09}. Based on this observation,
\citet{SB:Mannheim12} argued that a major part of the emission seen in
the TeV band from \ngc\ and M\,82 could be explained by the combined
emission from individual PWNe. The authors used the Galactic
population of TeV-detected PWNe to estimate the contribution of
leptonic \g-ray emission to the signal seen from both SB
galaxies. However, the environment in which most of these PWNe evolve
is significantly different from the physical conditions in SB
nuclei. In this work we follow the idea of \citet{SB:Mannheim12}, but
study the PWN population and the resulting non-thermal emission in the
SB based on the SFR and time-dependent pulsar injection power, rather
than the (incomplete) sample of Galactic TeV PWNe.

\section{PWN in a Starburst environment}\label{sec:ind}

Pulsar wind nebulae are formed when the cold ultra-relativistic wind
from a pulsar interacts with its environment. Particles in PWNe are
accelerated up to $\sim1$\,PeV and produce synchrotron and Inverse
Compton (IC) emission from radio to VHE \g-ray energies in
interactions with magnetic and radiation fields, respectively
\citep[e.g.][]{GaenslerSlane2006}. The majority of PWNe in the Milky
Way evolve in the Galactic disc, where the target radiation fields for
the IC process are the Cosmic Microwave Background (CMB), visible
starlight, and reprocessed starlight in the infrared. The typical
energy densities of these radiation fields are $U_{\mathrm{ph}} \sim
1$\,eV\,cm$^{-3}$. The dominant radiation field in the nuclear region
of SB galaxies, on the other hand, is far-infrared (FIR) radiation
with orders of magnitude higher energy density. \citet{NGC253:Melo}
for example found a total FIR luminosity of $2\times10^{10}\,L_\odot$
in the SB region of \ngc. The \ngc\ SB has a cylindrical shape with a
radius of $\sim150$\,pc and a half-height of $\sim60$\,pc
\citep{NGC253:Weaver02}. Assuming that each point in the cylinder
radiates with the same emissivity at the temperature of the cold dust
($\approx$50\,K), $U_{\mathrm{ph,NGC\,253}}\simeq2400$\,eV\,cm$^{-3}$,
a factor 2.9 higher than for a point source with the same luminosity
at 150\,pc distance \footnote{Note that \citet{Atoyan1996} derive a
  factor of 2.24 for the case of a spherical emission region.}. This
value is more than three orders of magnitude higher than is typical in
the ISM and implies that the cooling time of highly-relativistic
electrons in the SB region is much shorter, i.e. $\sim200$\,yrs
compared to $\sim 3\times 10^5$\,yrs for a 1\,TeV electron. IC cooling
in such an environment will dominate over synchrotron losses unless
$B\gtrsim 500\,\mu$G.

The central source that powers the PWN is a pulsar that spins down and
converts a fraction of its rotational energy into non-thermal
particles and subsequently into radiation. Over time the pulsar's
rotation slows down, and the energy put into the nebula decreases. The
spin-down luminosity evolves in time as

\begin{equation}\label{eq1}
  \dot{E}(t) = \dot{E}_0/(1+t/\tau)^p,~\mathrm{with}~\tau =
  P_0/\dot{P}_0(n-1),
\end{equation}
where $\tau$ is the characteristic spin-down time, given by the birth
period of the pulsar $P_0$ and the first derivative of $P_0$. $p = (n
+ 1)/(n - 1)$, where $n$ is the braking index and is measured only for
a handful of pulsars, where it lies between $\sim2$ and 3. In order to
investigate the effect of the different radiation fields we will apply
a single-zone, time-dependent model, where particles are injected
according to Equation~\ref{eq1}. This approach is similar to the one
followed in \citet{GC:Hinton07}.

\begin{figure*}
  \centering
  \resizebox{0.95\hsize}{!}{\includegraphics[]{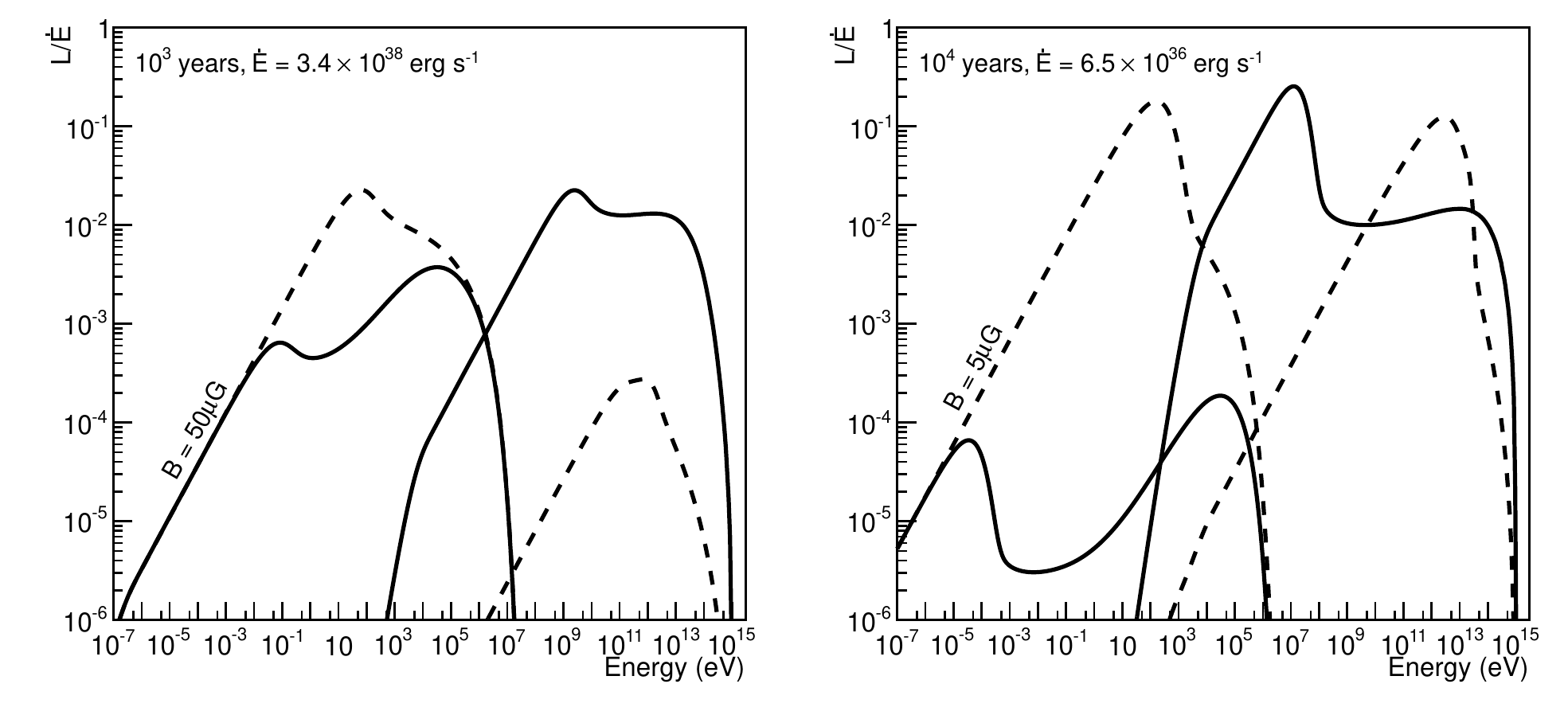}}
  \caption{Fraction of the total power injected in electrons that is
    radiated in different spectral bands for a PWN with a
    ``Crab-like'' birth period of $P_{0}=20$\,ms and a braking index
    $n=3$, after $10^3$\,yrs (left) and $10^4\,$yrs (right). The PWN
    is modeled as evolving in a SB-like (solid line) and ISM-like
    environment (dashed line), with a constant magnetic field strength
    of $50\mu$G (left) and $5\mu$G (right). The ISM radiation fields
    considered are, the CMB, a cold dust component with $T=50$\,K and
    a starlight component with $T=5000$\,K. Both radiation fields are
    assumed to have $U_{\mathrm{ph}}=1$\,eV\,cm$^{-3}$. The SB
    radiation field has a temperature of $T=50$\,K and
    $U_{\mathrm{ph}}=2400$\,eV\,cm$^{-3}$. Note the difference in
    current spin-down power $\dot{E}$ of a factor of $\sim50$ between
    10$^3$\,yrs and 10$^4$\,yrs. The dominance of IC emission in the
    SB environment at all times is clearly visible.}
\label{fig1}
\end{figure*}

Figure~\ref{fig1} shows the fraction of the total injected electron
power that is radiated in different spectral bands for a PWN evolving
in both SB-like, and ISM-like environments for different ages.
Particles are injected according to a power-law in energy $N(E)\propto
E^{-\alpha}$ with index $\alpha = 2.0$. Recent studies of prominent
Galactic PWNe \citep[e.g.]{PWN:Zhang08} suggest that the conversion
efficiency of spin-down power into non-thermal particles $\epsilon$ is
$\sim1/3$, with a recent lower limit of 0.3 derived for MSH~15--52
\citet{MSH:Schoeck10}. However, the fraction of spin-down power that
is converted into magnetic fields in young PWN is known to be small
\citep[e.g.]{PWN:Kennel1984a, PWN:delZanna06, PWN:Gelfand09,
  PWN:Tanaka10}, so that $\epsilon$ may approach 1. Non-radiative
losses and/or complex spectral shapes for the lower energy particles
may be responsible for the apparently lower injected power in
particles \citep[e.g.]{Spitkovsky08, Slane10}. In the following we
assume a constant value of $\epsilon=0.5$ and neglect non-radiative
losses, with an estimated uncertainty of a factor of 2 in either
direction.

Figure~\ref{fig1} illustrates that emission from an object evolving in
a typical ISM region is likely to be synchrotron-dominated at early
times, when the $B$-field is high. At later stages, when the magnetic
field decreases as the nebula expands, the fraction of energy
converted into IC emission is expected to increase. This effect
provides a plausible explanation for the number of bright, middle-aged
PWNe detected in VHE \g\ rays that have rather low X-ray flux
\citep[see e.g.][]{deJager09}. It has been suggested that such (more
numerous) middle-age PWN may be the main contributors to the VHE
emission of SB galaxies \citep{SB:Mannheim12}. However, the energy
output of a Crab-like pulsar after $10^3$ years is expected to be a
factor of $\sim50$ higher than at an age of $10^4$ years, more than
compensating for the smaller number of young objects. In a typical ISM
environment these younger objects would be expected to be
synchrotron-dominated, with low IC efficiency. Due to the high energy
density of radiation in a SB region, however, a large fraction of the
non-thermal energy is expected to be emitted as IC radiation at GeV --
TeV energies. Since the cooling time for the IC process is so short in
this environment, electrons with an energy as low as $\sim20$
(200)\,GeV are efficiently cooled for a PWNe age of $10^4 (10^3)$
years (Figure~\ref{fig1}, right).

\section{Emission from the PWN Population}\label{sec:pop}

The results obtained in the previous section suggest that the recent
particle injection history is of most importance when studying the HE
and VHE \g-ray emission from a population of PWNe in SB galaxies. In
the following we will investigate how a population of PWNe might
evolve in an environment similar to the one found in the SB nuclei of
\ngc\ and M\,82. The most important pulsar-related properties that
impact the overall level of non-thermal emission from PWNe in SB
galaxies are (1) the number of pulsars that form PWNe, (2) the
efficiency $\epsilon$ with which these convert rotational spin-down
energy to non-thermal particles, and (3) the distribution of pulsar
birth periods, as these determine the time evolution of particle
energy input.

\subsection{\ngc}

Based on the FIR luminosity in the SB nucleus of \ngc,
\citet{NGC253:Melo} derive a SFR of
$\sim3.5\,M_\odot$\,yr$^{-1}$. With a Salpeter initial mass function
and assuming that only stars with masses between $8\,M_\odot$ and
$20\,M_\odot$ undergo type-II SN explosions and hence form pulsars, we
estimate a type-II SN rate in the SB nucleus of \ngc\ of
0.02\,yr$^{-1}$. This value is close to the estimate by
\citet{NGC253:Engelbracht98} of 0.03\,yr$^{-1}$, but considerably
lower than the 0.08\,yr$^{-1}$ as estimated by \citet{vanBuren94}.
In the following we assume that all pulsars that are left behind by SN
type-II explosions form a PWN and evolve as discussed in
Section~\ref{sec:ind}. A key parameter for estimating the flux from
these systems is the typical birth period $P_{0}$ of pulsars, as the
energy available for particle acceleration is $\propto
P_{0}^{-2}$. Birth periods have been estimated for only a small number
of objects, for example for the Crab pulsar where $P_0 \approx
16\,$ms.  However, a relatively recent study based on the population
of radio pulsars suggests a mean birth period of $\langle P_0 \rangle
\simeq 300$\,ms, with a spread of $\sigma_{P_0}\simeq150$\,ms
\citep{Faucher06}. More recently, a model taking into account the
population of \g-ray pulsars as detected by the Fermi-LAT suggests a
revision of the mean pulsar birth period down to $\langle P_0\rangle
\simeq 50$\,ms \citep{Watters11}. The Crab may therefore represent a
rather extreme case. In the following we assume a fixed value of
$P_{0}=35$\,ms and discuss the implications of this choice below. A
conversion efficiency $\epsilon = 0.5$ is chosen as discussed
above. The magnetic field in a PWN is expected to decrease with time,
however, in the SB environment energy losses are likely dominated by
IC emission at all times and the $B$-field evolution is not expected
to result in a modified particle energy distribution. We therefore
assume no magnetic field evolution in the nebulae and use a mean
$B_{\mathrm{PWN}}$ field value of 25\,$\mu$G. As in the previous
section we assume $n=3$, and injection of particles with power-law
index $\alpha=2.0$.

\begin{figure*}
  \centering
  \resizebox{0.475\hsize}{!}{\includegraphics[]{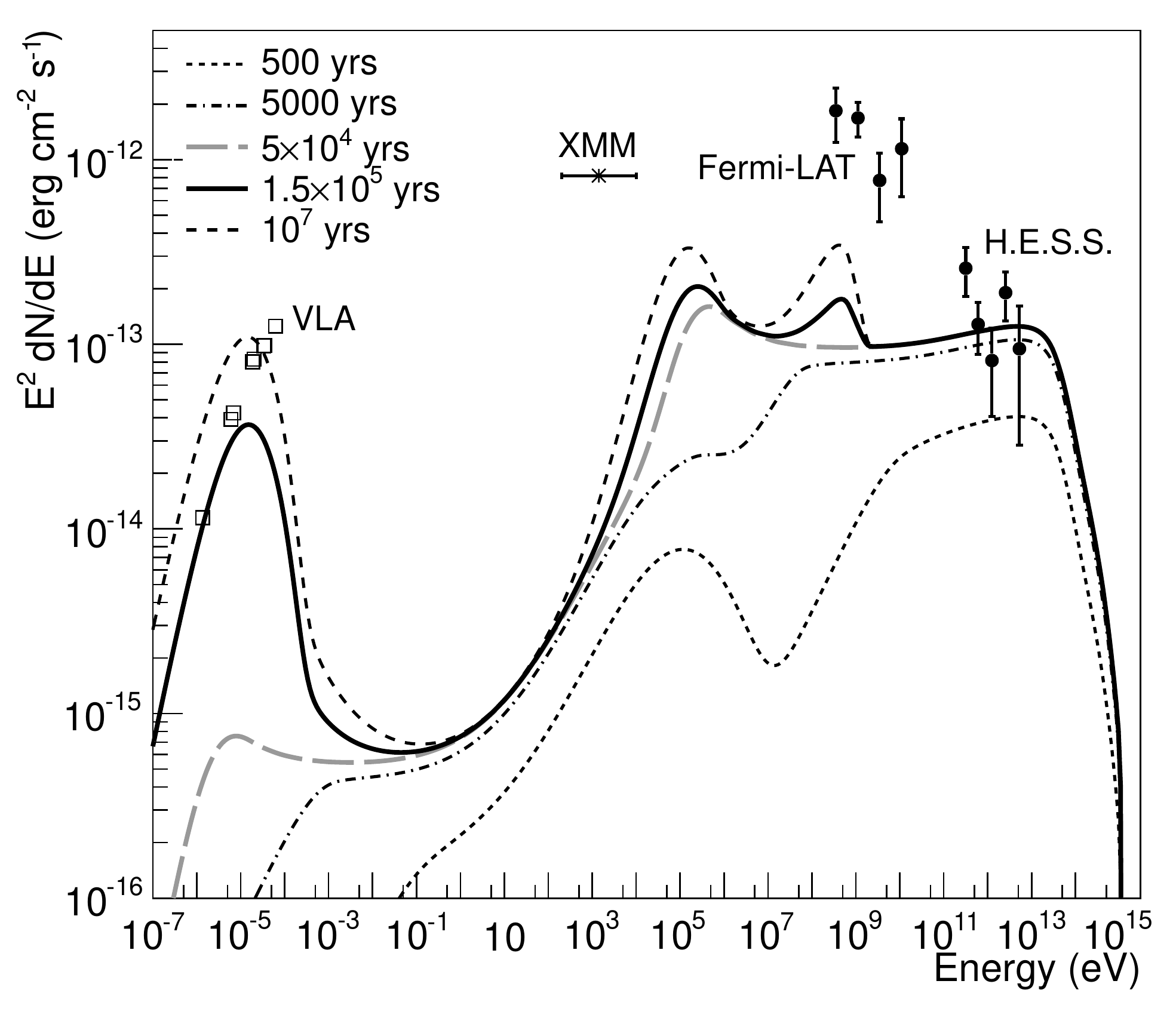}}
  \resizebox{0.475\hsize}{!}{\includegraphics[]{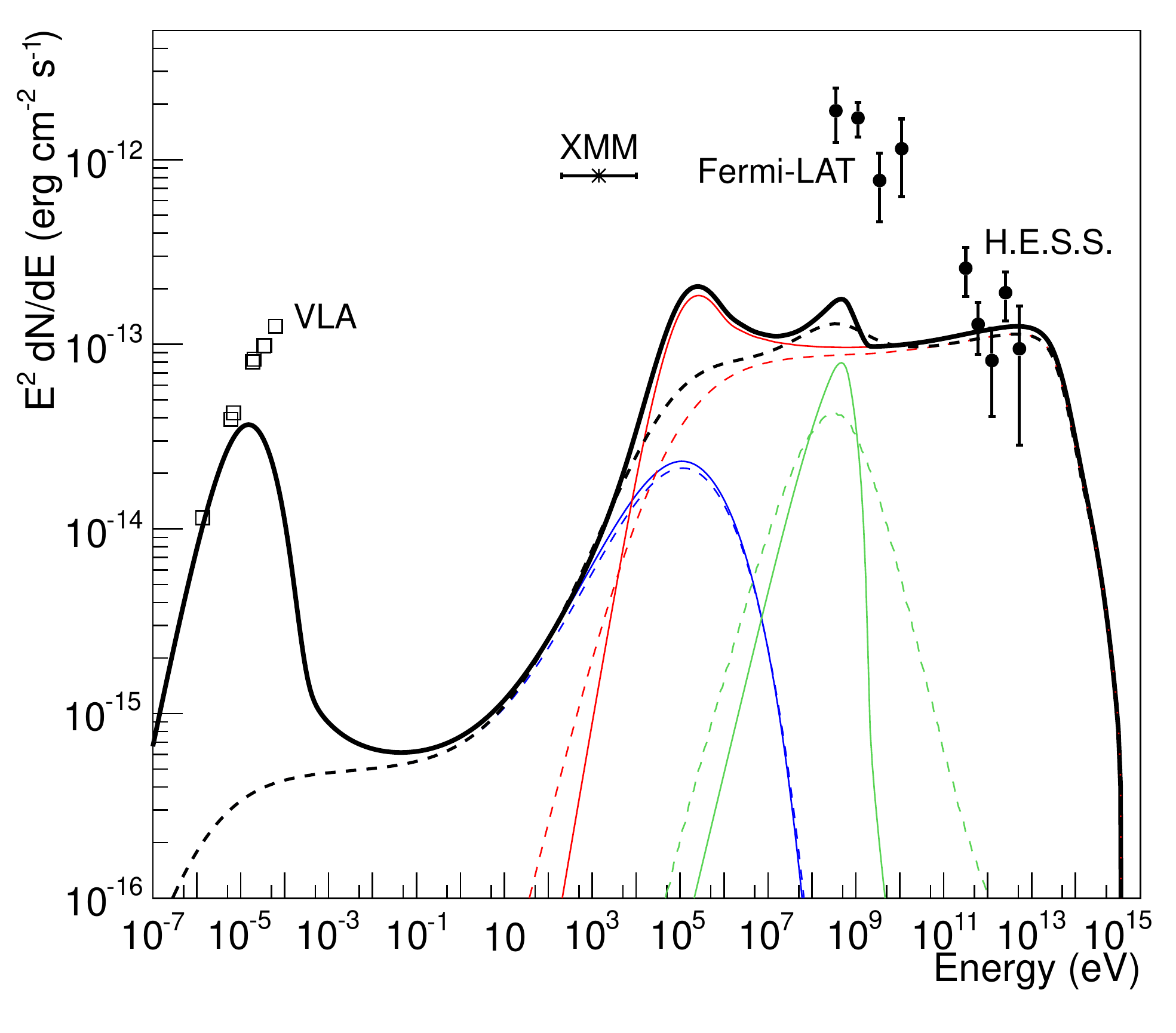}}
  \caption{Predicted SEDs for PWNe populations in the SB environment
    of \ngc\ at different evolutionary stages. Also shown is radio
    data from the Very Large Array \citep[VLA, ][]{Carilli96}, the
    X-ray point is emission from the central source X\,34 as given by
    \citet{NGC253:Pietsch01} and the Fermi-LAT and H.E.S.S. data is
    from \citet{HESS:NGC253_2012}. {\it Left:} The integrated emission
    from {\it individual} PWNe is indicated by the long-dashed, grey
    curve. The equilibrium spectrum for all objects with ages less
    than $1.5\times10^5$\,yrs is given by the solid black curve. The
    effect of particle confinement in the SB region over the age of
    the SB episode of $10^7\,$yrs is indicated by the short-dashed
    black curve. {\it Right:} Equilibrium spectrum of \ngc\ for the
    two-zone, time-dependent model (straight lines) and for the simple
    one-zone model (dashed lines), both of which are described in the
    main text. The different components of the emission are IC (red),
    Bremsstrahlung (green) and synchrotron (blue).}
\label{fig2}
\end{figure*}

The left-hand panel of Figure~\ref{fig2} shows a model spectral energy
distribution (SED) for the total non-thermal emission of a population
of PWNe in \ngc's SB at different evolutionary stages. This
calculation is a two-zone, time-dependent model, where particles are
injected in each PWN according to Equation~\ref{eq1}. The number of
PWN of a given age is calculated in 0.25 dex steps under the
assumption of a constant pulsar birth rate as given above. Injected
particles cool via IC and synchrotron processes in the PWN
environment. For objects older than $5\times10^4$ years, we assume
that no more injection of particles in the PWN occurs, that all
particles escape and continue to cool in the SB region of
\ngc. Non-thermal emission generated by this population of escaped
particles add to the {\it diffuse} emission of \ngc's SB region. As
the SB nucleus is characterised by a much higher magnetic field and
average density ($B_{\mathrm{SB}}\simeq 250\,\mu$G, $n_H \simeq
250\,$cm$^{-3}$) synchrotron, Bremsstrahlung and Coulomb losses become
important for particles that have escaped the PWNe. The high thermal
and cosmic-ray pressure in the SB region leads to the formation of a
SB wind which effectively removes material from the centre of the
galaxy. The height of the SB region of $H\simeq60$\,pc and the wind
speed of $v_{\mathrm{wind}}\simeq300$\,km\,s$^{-1}$
\citep{NGC253:Voelk06} implies an advective loss time of
$\tau_{\mathrm{ad}} = (H/2)/v_{\mathrm{wind}}\simeq 10^5$\,yrs, after
which particles are removed from the SB region. Hence, the SED for an
age of $1.5\times10^5$\,yrs may well be representative of the
equilibrium spectrum produced by all PWNe over the past 150\,kyr. The
SED for an age of $5\times10^4$\,yrs, on the other hand, shows the
contribution of non-thermal particles that have not yet escaped their
PWNe. This SED therefore represents the contribution of {\it
  individual} PWN emission (rather than diffuse emission) to the
overall non-thermal emission from the SB nucleus of \ngc.

The right-hand panel of Figure~\ref{fig2} illustrates how the
different energy-loss mechanisms shape the resulting SEDs. In the
employed model, IC emission dominates at photon energies above
$\sim100$\,keV due to efficient cooling of non-thermal particles in
the strong omnipresent FIR radiation field. The peak at $\sim500$\,GeV
in the \g-ray spectrum is caused by particles that have left their
low-density PWNe environments after $5\times10^4$\,yrs and emit
Bremsstrahlung in the dense gas of the SB region. The synchrotron
emission spectrum is complex with a double-hump structure. Since the
magnetic field in the SB region is so high, electrons with an energy
of $\gtrsim1$\,GeV that have left the PWNe predominantly lose energy
via the synchrotron process and contribute to the radio emission. The
second hump of the synchrotron spectrum in the X-ray domain is caused
by the population of higher-energy particles that cool in the magnetic
field of young PWNe. This figure also compares the two-zone model
described above to a simple one-zone model. In this approach,
particles are injected continuously rather than according to
Equation~\ref{eq1}, representing the semi-continuous nature of
particle injection from the population of PWN as a whole. The
environment in which particles cool in this calculation has been
chosen to match the PWN environment for the target fields for
processes that are important for the highest energy particles
($B\simeq 25\,\mu$G and SB radiation field), but using the mean SB
density ($n_H \simeq 250\,$cm$^{-3}$) for the calculation of
Bremsstrahlung, as particles are expected to escape their PWN on
roughly the Bremsstrahlung cooling time in the SB. Although the IC and
Bremsstrahlung parts of the \g-ray spectra are quite similar, the
synchrotron components look very different. This is a direct
consequence of the two-zone model, with the low (high) magnetic field
strengths inside (outside) of PWNe.

\subsection{M\,82}

In the following we apply the same two-zone model calculation to M\,82
and compare it to multi-wavelength data from radio to VHE. The FIR
emission from M\,82's SB nucleus can be well described by a dominant
dust component with a temperature of $\sim 50$\,K and a luminosity of
$3.8\times 10^{10}\,L_{\odot}$ \citep{M82:Colbert99}. The inferred SFR
is $\sim10\,M_\odot$\,yr$^{-1}$, considerably higher than in \ngc, and
implies a type-II SN rate of $\sim$0.06\,yr$^{-1}$. Adopting the
disc-like SB geometry of 150\,pc radius and 60\,pc height as used in
\citet{Strickland00}, the higher intensity of the FIR radiation in
M\,82's SB region also leads to a higher radiation field energy
density of $U_{\mathrm{ph,M82}}\simeq5000$\,eV\,cm$^{-3}$. The
magnetic field adopted here is the same as in \ngc's SB:
$B_{\mathrm{SB}}=250\mu$G, consistent with estimates of $200\mu$G --
$400\mu$G \citep{Thompson06,Strickland07}. For the average density we
again assume a value of $n_H \simeq 250\,$cm$^{-3}$.

\begin{figure}
  \centering
  \resizebox{0.95\hsize}{!}{\includegraphics[]{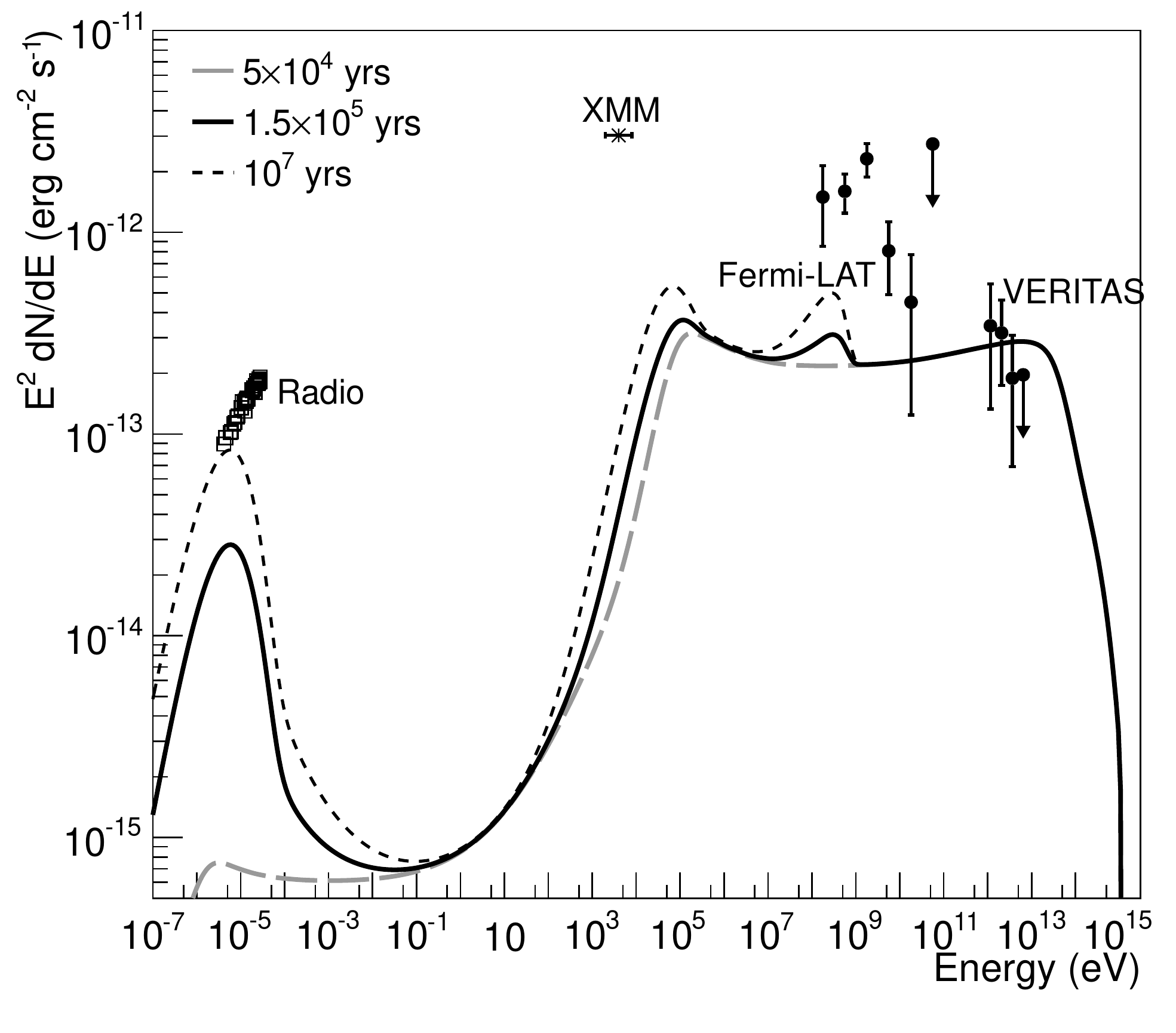}}
  \caption{As Figure~\ref{fig2}, but for the emission from PWNe that
    evolve in the SB environment of M\,82. Multi-wavelength data is
    from \citet[][radio]{M82:Colbert99},
    \citet[][X-ray]{Strickland07}, \citet[][HE \g\ rays]{Fermi:SB} and
    \citet[][VHE \g\ rays]{VERITAS:M82}.  }
  \label{fig3}
\end{figure}

Figure~\ref{fig3} shows the result for the two-zone model as described
above. Particles are injected according to Equation~\ref{eq1} and for
the estimated SN rate of $0.06\,$yr$^{-1}$. In order not to
overproduce the emission measured by VERITAS, the pulsar birth period
has been reduced to 40\,ms. All other pulsar-related parameters are
kept as above. As before the time after which particles leave the PWNe
into the surrounding SB environment is fixed to $5\times10^4$
years. Particles are again assumed to leave the SB region via
energy-independent transport in the SB wind after
$1.5\times10^5$\,yrs. The calculated SEDs have very similar shape to
the curves obtained for \ngc, as expected due to the similar target
radiation field, magnetic fields, and average particle densities in
the SB regions.

\section{Discussion}\label{sec:discussion}

Figures~\ref{fig2} and \ref{fig3} show that with a suitable choice of
model parameters the VHE \g-ray emission of \ngc\ and M\,82 may be
dominated by PWN as suggested by \citet{SB:Mannheim12}. In the
following we discuss the plausibility of these parameter values.

{\bf Magnetic and radiation fields. } Figure~\ref{fig2} shows that
young PWNe are expected to dominate the HE and VHE \g-ray
emission. Magnetic field strengths in young objects such as the Crab
Nebula are estimated to be $\sim100-400\mu$G \citep{Atoyan1996,
  PWN:Zhang08, PWN:Volpi08, PWN:Tanaka10}. The precise strengths of
magnetic fields in these systems has little effect on the VHE fluxes
unless the magnetic energy density $U_{B}$ exceeds $U_{\rm ph}$. Note
PWNe that dominate in the \citet{SB:Mannheim12} model have much lower
magnetic fields \citep{Okkie09}. Our SB energy density estimates are
based on a simplified geometry but are relatively robust, and exceed
$U_{B}$ as long as $B<500\,\mu$G (or even higher for M\,82). IC
dominance over (at least most of) the early lifetime of pulsars in the
SBs therefore seems likely.

{\bf Particle Escape.} In our model we assume that all particles
escape the PWNe after $5\times10^4$\,yrs. Smaller values of the escape
time, and energy-dependent escape are certainly plausible \citep[see
e.g.][]{Vela:Hinton11} but have modest impact on the resulting SED
except to boost diffuse synchrotron emission at high radio frequencies
and bremsstrahlung emission at $\sim1$ GeV. The escape time of
particles from the SB region must be less than $\sim1$ Myr to avoid
violating constraints on the diffuse radio emission. However, energy
independent escape via the SB wind on much shorter timescales seems
likely \citep{NGC253:Voelk06}. For even higher wind speeds
\citep[e.g.][]{M82:Strickland09}, particles would escape the SB region
faster and the Bremsstrahlung and radio components would be lower.

{\bf Birth Period.} Average values of $P_{0}$ (or strictly
$(<P_{0}^{2}>)^{1/2}$) of $\sim$35\,ms and $\sim$40\,ms are required
to match the VHE fluxes of \ngc\ and M\,82, respectively. Whilst these
values are somewhat lower than typical assumptions for pulsar
populations, it is certainly possible that these values are
representative of recently formed pulsars in these
objects. Furthermore, a plausible increase of the injection efficiency
$\epsilon$ and/or a systematic underestimate in the SN rate, would
increase these values into a more comfortable range.

{\bf Injection spectrum.} The assumed hard injection spectrum
($\alpha=2$) maximises the VHE flux. Softer spectra at injection are
plausible, and would be consistent with the measured VHE spectra, but
increase the required energy input dramatically. For example for
$\alpha=2.3$ the required birth periods are $\sim$15\,ms and fast
particle escape ($t_{\rm esc}\ll10^{5}$ years) is required for
consistency with radio limits. As a consequence PWN cannot explain the
emission detected with Fermi from these objects.

\section{Conclusions and Outlook}

Considering the very different evolution of particle populations in
starburst environments to typical Galactic environments we
nevertheless reach a similar conclusion to that of
\citet{SB:Mannheim12}: that PWN can plausibly contribute to the TeV
emission of \ngc\ and M\,82. The Fermi emission cannot be explained in
this framework on energetics grounds and the standard explanation of
diffuse emission from strong interactions of accelerated cosmic rays
remains the most likely possibility. As a cut-off in this component
just before the TeV range seems unlikely, the observed VHE emission
may represent a superposition of a harder PWN component on the diffuse
p-p emission. CTA \citep{CTA:Actis11} will have the sensitivity to
search for the spectral feature associated with PWN and place tight
constraints on the pulsar population.

\section*{Acknowledgements}
S.O. acknowledges the support of the Humboldt foundation by a
Feodor-Lynen research fellowship.

\bibliographystyle{mn2e_williams}
\bibliography{NGC253_PWN}
\label{lastpage}

\end{document}